\documentclass{aipproc}

\layoutstyle{6x9}

\usepackage{graphicx}
\usepackage{colordvi}
\usepackage{epsf}
\usepackage{amsmath}

\begin{document}
\title{Quark model study of the semileptonic  $B\to\pi$ decay}


\author{C. Albertus}{ address={School of Physics and Astronomy,
  University of Southampton, Southampton SO17 1BJ, United Kingdom.}  }

\author{J. M. Flynn}{ address={School of Physics and Astronomy,
  University of Southampton, Southampton SO17 1BJ, United Kingdom.}  }

\author{E. Hern\'andez}{ address={Grupo de F\'\i sica Nuclear,
  Departamento de F\'\i sica Fundamental e IUFFyM, Universidad de
  Salamacanca, E-37008 Salamanca, Spain.}}

\author{J. Nieves}{address={Departamento de F\'\i sica At\'omica,
Molecular y Nuclear, Universidad de Granada, E-18071 Granada, Spain.}}

\author{J. M. Verde--Velasco}{ address={Grupo de F\'\i sica Nuclear,
  Departamento de F\'\i sica Fundamental e IUFFyM, Universidad de
  Salamacanca, E-37008 Salamanca, Spain.}}

\begin{abstract}
The semileptonic decay $B\to\pi l \bar\nu_l$ is studied starting from
a simple quark model and taking into account the effect of the $B^*$
resonance.  A novel, multiply subtracted, Omn\`es dispersion relation
has been implemented to extend the predictions of the quark model to
all physical $q^2$ values. We find
$|V_{ub}|=0.0034\pm0.0003{\rm(exp.)}\pm0.0007{\rm(theory)}$, in good
agreement with experiment.

\end{abstract}

\maketitle
\section{Introduction}

The measurement of the exclusive semileptonic decay $B\to\pi
l\bar\nu_l$ can be used to determine de Cabibbo-Kobayashi-Maskawa
(CKM) matrix element $|V_{ub}|$.  With no flavor symmetry constraining
the hadronic matrix elements, the errors on $|V_{ub}|$ are currently
dominated by theoretical uncertainties, being a determination of
$|V_{ub}|$, with well understood uncertainties, a priority of heavy
flavor physics.  The application of Watson's theorem to the $B\to\pi$
semileptonic decay allows one to write a dispersion relation for each
of the form factors entering in the hadronic matrix element. This
leads to the so-called Omn\`es representation, which can be used to
constrain the $q^2$ dependence of the form factors assuming some
knowledge of the elastic $\pi B\to \pi B$ scattering amplitudes. The
use of multiple subtractions will allow to combine predictions from
various methods in different $q^2$ regions.  In this talk, we show how
this Omn\`es scheme can be used to combine the results at $q^2=0$
of the relevant hadron $B\to\pi l\bar \nu_l$ form factors from light
cone sum rules (LCSR) calculations, with those obtained from a simple
nonrelativistic constituent quark model (NRCQM) in its region of
applicability, near zero recoil. In this way we end up, with an
accurate description of the differential decay rate, except for
$V_{ub}$, in the whole physically accessible $q^2$ range.  We also use
a Monte Carlo simulation to estimate the theoretical error bands of
our procedure.

\section{NRCQM: Valence quark, $B^*$ resonance and 
Omn\`es representation}

The hadronic matrix element for $B^0\to\pi^-l^+\nu_l$ can be parametrized in
terms of two dimensionless form factors, $f^+$ and $f^0$,  of which only 
$f^+$ contributes  for massless final leptons.
\begin{figure}
%\begin{center}
%\rotatebox[origin = lb]{90}{\scalebox{3}[1]{preliminary fig.}}
\includegraphics[clip,scale=0.6, bb= 250 500 520 800]{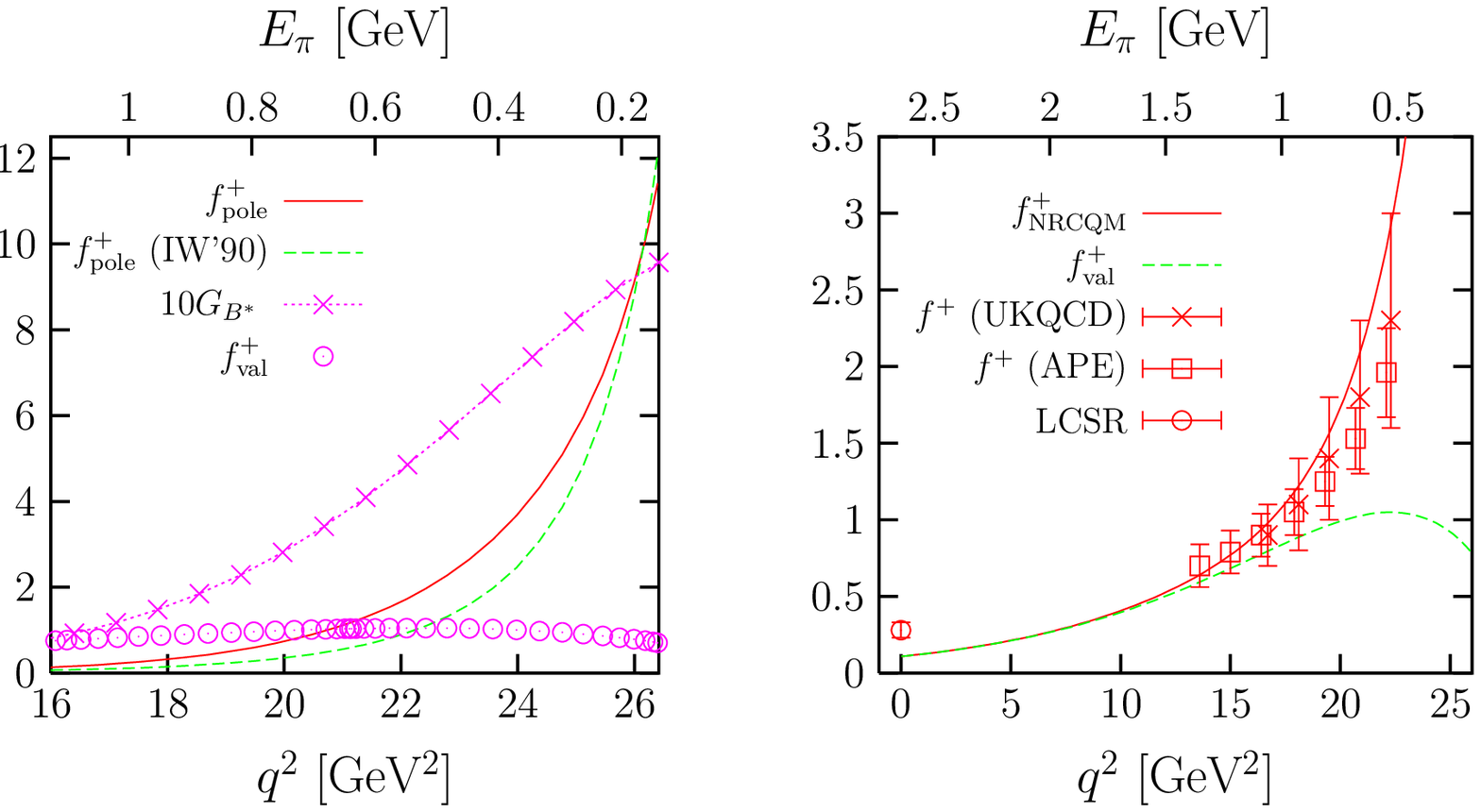}
\caption{\label{fig1}Valence quark (val), and valence quark plus $B^*$
contribution (NRCQM) to $f^+$. We also plot lattice QCD and LCSR $f^+$
results. See Ref.~\cite{1} for details.}
%\end{center}
\end{figure}
In Figure~\ref{fig1} we show under ``$val$'' the NRCQM prediction for
$f^+$ when considering only the valence quark contribution. The
description fails in the whole $q^2$ range: from the region close to
$q^2_{\rm max}$, where a nonrelativistic model should work best, to
the opposite end where the pion is ultrarelativistic and thus
predictions from a nonrelativistic scheme are unreliable. As first
pointed out in Ref.~\cite{2}, near zero recoil the $B\to\pi l^+\nu_l$
decay is dominated by the effects of the $B^*$ resonance, which is
quite close to $q^2_{\rm max}$.  These effects of the $B^*$ resonance
must be added as a distinct coherent contribution. We have
consistently evaluated within our model the $B^*$ resonance
contribution to the $f^+$ form factor (See Ref.~\cite{1} and
references therein for details). The result is that the $B^*$
resonance plays a role only near $q^2_{max}$, being strongly
suppressed by a soft hadronic vertex outside that region. The
inclusion of the $B^*$ resonance contribution to the form factor
($f^+_{NRCQM}$ in the figure) improves the simple valence quark
contribution down to values around $15$ GeV$^2$. Below that the
description is still poor.

%
%\subsection{Omn\`es representation}
%
We have now used the Omn\`es representation to combine the NRCQM
predictions at high $q^2$ with the LCSR result at $q^2=0$. This
representation requires as an input the elastic $\pi B\to \pi B$ phase
shift $\delta(s)$ in the $J^P=1^-$ and isospin $I=1/2$ channel plus
the form factor at different $q^2$ values below the $\pi B$ threshold
where we will perform the subtractions.  With a large enough number of
subtractions only the phase shift at or near threshold is needed. We
can then approximate $\delta(s)\approx\pi$ (Levinson's theorem) which
renders our calculation analytic.

\begin{figure}
%\begin{center}
\resizebox{0.65\columnwidth}{!}{\includegraphics{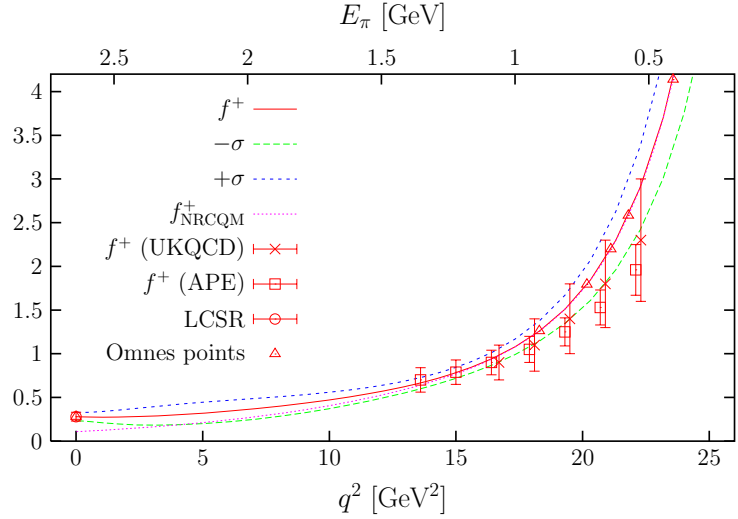}}
\caption{\label{fig3}The solid line represents the
Omn\`es improved form factor. The subtraction points are denoted by
triangles. The $\pm\sigma$ lines show the theoretical uncertainty band
on the Omn\`es form factor. We compare with previous lattice results.}
%\end{center}
\end{figure}
In Figure~\ref{fig3} we show with a solid line the form factor
obtained using the Omn\`es representation. We have used as subtraction
points five $q^2$ values between $18$ GeV$^2$ and $q^2_{\rm max}$, for
which the NRCQM predictions (valence+pole) have been used, plus the
LCSR prediction at $q^2=0$. We have paid special attention to the
estimation of theoretical uncertainties that come from two main
sources: (i) uncertainties in the quark--antiquark nonrelativistic
interaction and (ii) uncertainties on the $[g_{B^*B\pi}f_{B^*}]$
product and on the input to the multiply subtracted Omn\`es
representation.  As a result, we obtain the $68\% $ confidence level
region enclosed between $\pm\sigma$ lines. See Ref.~\cite{1} for
details. Comparing with the experimental decay width, we obtain
$|V_{ub}|=0.0034\pm 0.0003({\rm exp.}) \pm 0.0007({\rm theo.})$ in
good agreement with a recent experimental determination by the CLEO
Collaboration~\cite{4}.

\section{Acknowledgments}
This work was supported by DGI and FEDER funds, under Contracts
No. FIS2005-00810, BFM2003-00856 and FPA2004-05616, by the
Junta de Andaluc\'\i a and Junta de Castilla y Le\'on under Contracts
No. FQM0225 and No. SA104/04, and it is a part of the EU integrated
infrastructure initiative Hadron Physics Project under Contract
No. RII3-CT-2004-506078. J.M.V.-V. acknowledges an E.P.I.F contract
with the University of Salamanca. C. A. acknowledges a research
contract with the University of Granada.

\end{document}